\documentclass[twocolumn,prl,aps,showpacs]{revtex4}
\usepackage{amssymb}
\usepackage{amsmath}
\usepackage{graphicx}
\usepackage{dcolumn}
\usepackage{bm}

\begin{document}

\preprint{Contact v.4}
\title{Spin extraction theory and its relevance to spintronics}
\author{H. Dery} \email{hdery@ucsd.edu}
\author{L. J. Sham}
\affiliation{Department of Physics, University of California San
Diego, La Jolla, California, 92093-0319}
\begin{abstract}
Extraction of electrons from a semiconductor to a ferromagnet as
well as the case of injection in the reverse direction may be
formulated as a scattering theory. However, the presence of bound
states at the interface arising out of doping on the semiconductor
side must be taken into account in the scattering theory.  Inclusion
of the interface states yields an explanation of a recent result of
spin imaging measurement which contradicts the current understanding
of spin extraction. The importance of an extraction theory to
spintronics is illustrated by an application to a spin switch.

\end{abstract}
\maketitle


Spin injection experiments in biased Fe/GaAs structures show that
the net spin of injected electrons from the ferromagnet to the
semiconductor is parallel to the majority spin population of the
ferromagnet \cite{Hanbicki_APL03,Crooker_Science05,Lou_PRL06}. This
implies that more majority than minority spins cross the junction.
If extraction of spins \textit{follows the same scattering process}
as injection but in the other direction, then by conservation of
spins the paramagnetic semiconductor is left with spin accumulation
parallel to minority spins. However, a recent experiment by Crooker
\textit{et al.} \cite{Crooker_Science05} had shown that in both
injection and extraction the same spin species dominates the
accumulation in the semiconductor. In these experiments the heavily
doped profile at the interface, needed for the creation of thin
Schottky barriers, extends over $\simeq$30~nm. On the other hand,
the underlying bulk semiconductor is lightly doped. This
inhomogeneous doping localizes electrons in surface bands next to
the Schottky barrier. In this letter we study the coupling of these
electrons to the ferromagnet. When this coupling dominates the
transport, the spin accumulation matches the experimental
observation.


\begin{figure} [!h]
\includegraphics[height=7cm,width=7.5cm]{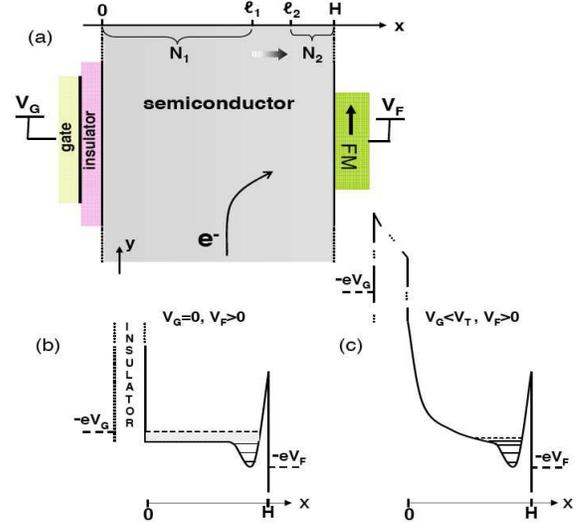} 
\caption{(a) A scheme of a spin switch. Electrons in the
semiconductor are drifted in the y direction until being redirected
into the ferromagnet (V$_F$$>$0). These electrons do not necessarily
originate from a spin-polarized source (zero bias reference). The
current path does not involve the gate due to the resistive
insulator. The semiconductor conduction band between the insulator
and the ferromagnet is shown schematically for two different gate
biases. In (b), the ferromagnet extract free electrons in addition
to a smaller part by localized electrons from the potential well.
This well is generated by the doping profile. In (c), the gate
voltage is negative enough to deplete the free electrons between the
insulator and the well. The extraction is dominated by tunneling of
localized electrons. These two cases result in opposite spin
polarity in the semiconductor region outside the sandwiched
structure.} \label{Fig:rev_fwd}
\end{figure}

Understanding of spin injection and extraction may have application
potential as illustrated by an electrically controlled spin switch,
shown in Fig.~\ref{Fig:rev_fwd}a. The gate regulates the density of
free electrons that can tunnel from the semiconductor to the
ferromagnet. If tunneling from the semiconductor free (localized)
electrons contributes the most to the current then the ferromagnet
favors extraction of spin-up (spin-down) electrons and a surplus of
spin-down (spin-up) electrons is left in the semiconductor. This
assumes a ferromagnet whose Fermi wave vector for spin-up electrons
is larger than for spin-down electrons,
$k_{m,\uparrow}$$>$$k_{m,\downarrow}$.
Figs.~\ref{Fig:rev_fwd}b~and~\ref{Fig:rev_fwd}c show characteristic
conduction band profiles in two gate bias regimes.

The spin-related antipodal behavior of free and localized electrons
may be understood with the help of $|t_{sc \rightarrow m}|^2$: the
square amplitude of a transmitted wave due to an incident plane wave
from the semiconductor side. The transmitted current of free
electrons is proportional to $k_{m}$$\cdot$$|t_{sc \rightarrow
m}|^2$. It is easy to show that $|t_{sc \rightarrow m}|^2$ decreases
with $k_m$. However, this dependance is weaker then $k_m^{-1}$ if
the Schottky barrier is relatively high and if the effective mass of
the semiconductor electron is noticeably smaller than that of the
narrow d-bands in the ferromagnet. Thus, the transmitted current of
free electrons increases with $k_m$. For localized electrons, the
transmitted current due to escape into the ferromagnet scales with
the decay rate of the bounded wave function. The conservation of
total reflection and transmission is therefore irrelevant, and the
current has a $k_m$ dependence which is somewhat similar to that of
$|t_{sc \rightarrow m}|^2$ rather than of $k_m$$\cdot$$|t_{sc
\rightarrow m}|^2$. The enhanced escape rate with decreasing $k_m$
is also compatible with the description of an alpha-particle decay
\cite{kemble_book,Gamow}.

The proposed switch follows the physics of spin extraction. Using
the labels in Fig.~\ref{Fig:rev_fwd}a, the doping is set at N$_1$ up
to a distance of $\ell_1$ from the insulator. This is followed by a
steep increment from N$_1$ to N$_2$ within a distance of
$\ell_2-\ell_1$ and the doping is kept at N$_2$ up to the
semiconductor/ferromagnet (S/F) interface in x=H. We summarize the
conditions needed for spin switching:
\begin{eqnarray}
(\text{A1}) &&  \!\!\! H\!\!-\!\ell_2 \!\approx
\!\sqrt{\frac{\epsilon_{r}\phi_b}{2\pi e^2 N_2}}, \qquad (\text{B1})
\,\,\, \ell_1 \!\! \approx \!\!\sqrt{\!\!\frac{\epsilon_{r}E_g}{2\pi e^2 N_1}}, \nonumber \\
(\text{A2}) && \!\!\! N_2\gg N_1, \qquad \qquad \qquad
(\text{B2})\,\,\, \ell_1 > \ell_{mfp}, \nonumber
\\ (\text{A3}) && \!\!\! \ell_2-\ell_1
\sim r_{_B},  \qquad \qquad \,\,\,\, (\text{B3}) \,\,\, N_1 \gg
N_{\alpha}. \nonumber
 \label{eq:current_density}
\end{eqnarray}
$\epsilon_{r}$ and $\phi_b$ are, respectively, the relative
permittivity and the built-in potential of the S/F contact. $E_g$
and $r_{_B}$ are, respectively, the band gap energy and the
electron's de-Broglie wavelength. $\ell_{mfp}$ is the electron mean
free path in the lower doping region. N$_\alpha$ is a characteristic
density at which the impurity band is merged into the conduction
band for low temperatures (and the chemical potential lies at the
vicinity of the conduction band edge). The conditions A1-A3 are
needed for the creation of surface bands next to the narrow Schottky
barrier which according to A1 extends in x$\epsilon$[$\ell_2$,H].
Condition A2 guarantees an excess of electrons in
x$\epsilon$[$\ell_1$,$\ell_2$] compared with the electron density in
x$\epsilon$[0,$\ell_1$]. These electrons are localized around the
dense ionized donors in this narrow region (A3). The complementary
conditions are needed for switching between extraction mechanisms.
When $V_G$$<$0, a depletion region is formed next to the insulator.
The threshold voltage, V$_G$=V$_T$, is defined when the conduction
band is bent by E$_g$ and the depletion region reaches its intrinsic
maximal width \cite{footnote0}. Condition B1 guarantees the
depletion region can reach the potential well, as shown in
Fig.~\ref{Fig:rev_fwd}c. Thus, the motion in the x direction is
quantized even for electrons whose energies are at the semiconductor
chemical potential. On the other hand, when V$_G$=0, as shown in
Fig.~\ref{Fig:rev_fwd}b, the ionized donors in
x$\epsilon$[0,$\ell_1$] are neutralized by \textit{free} electrons
(B2) and their rather high density guarantees that these free
electrons dominate the transport (B3). Unless otherwise mentioned,
we use the following n-type GaAs layer ($r_b$$\approx$10~nm,
$\epsilon_{r}$$\approx$12.6): $\ell_1$=75~nm, $\ell_2$=90~nm,
H=105~nm, N$_1$=4$\cdot$10$^{17}$~cm$^{-3}$ and
N$_2$=5$\cdot$10$^{18}$~cm$^{-3}$. At low temperatures,
$E_g$$\approx$1.5~eV, l$_{mfp}$$\sim$50~nm \cite{Konemann_PRB01} and
N$_\alpha$$\sim$2$\cdot$10$^{16}$~cm$^{-3}$.

\begin{figure}
\includegraphics[height=7cm,width=8cm]{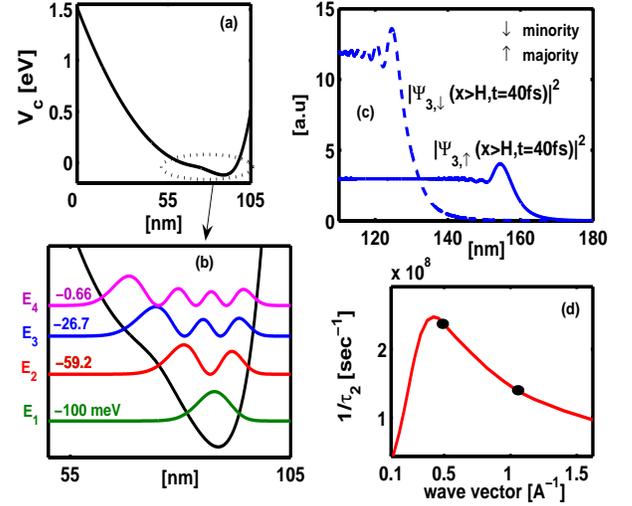}
\caption{(a) The conduction band potential in the semiconductor for
V$_f$=0.2~V and V$_G$=V$_T$. (b) Bound states at the bottom of the
band whose energies are below the semiconductor chemical potential
(zero level). The total electron areal densities are
n$_{1\rightarrow4}$=2.77, 1.66, 0.75, 0.03 [10$^{12}$~cm$^{-2}$].
(c) Spin dependent amplitude of the penetrated wave function in the
Fe side (x$>$H=105~nm) after 40~fs. The initial condition (t=0) for
both spins is the third bound state in the semiconductor side and
zero in the metal side. The enhanced group velocity of majority
electrons is the reason for their advanced wave front in the shorter
time. (d) Escape rate from the second bound state versus the
electron wave vector in a general metal case. The upper (lower)
marked dot refers to the case of minority (majority) electrons in
Fe.} \label{Fig:self_consistent_TDSE}
\end{figure}

The transport in the semiconductor is governed by the shape of the
conduction band. Fig.~\ref{Fig:self_consistent_TDSE}a shows the case
for V$_G$=V$_T$ as calculated by a self consistent scheme of the
Schrodinger-Poisson equations. The temperature is 10K, the bias
across the S/F junction is V$_F$=0.2~V, the built-in potential is
$\phi_{b}$=0.7~eV and the GaAs electron mass is $m_{sc}$=0.067m$_0$.
The free electrons are depleted due to the biased gate and the
localized electrons are found in one of 4 bound states shown in
Fig.~\ref{Fig:self_consistent_TDSE}b. The escape process is studied
in the following way. Initially, the wave functions are identical
for s=$\uparrow$ and s=$\downarrow$ and are taken as the i$_{th}$
bound state. In the metal side it is assigned with
$\psi_{i,s}$(x$>$H,t=0)=0. The time dependent Schrodinger equation
is numerically solved with the potential in the semiconductor side
being the self-consistent solution shown in
Fig.~\ref{Fig:self_consistent_TDSE}a. The potential in the metal
side is a simplified Fe model with free-electron mass and with
$k_{m,\uparrow}$=1.1~\AA$^{-1}$ ($k_{m,\downarrow}$=0.42~\AA$^{-1}$)
for majority (minority) electrons \cite{Slonczewski_PRB89}.
Fig.~\ref{Fig:self_consistent_TDSE}c shows the third quasi-bound
state penetrated wave function in the metal side after 40~fs. Note
that electrons of the Fe minority species have bigger penetrated
amplitude. This behavior persists for longer times and for all bound
states. For the calculations we have used a one dimensional box
x$\epsilon$[0,L$_b$=200~nm] with discrete transparent boundary
conditions to prevent reflections from the edges \cite{Arnold_2003}.
The ``leakage'' of the wave functions into the ferromagnet is slow
due to the S/F Schottky barrier. The escape rate is practically
constant in time (an exponential decay process) and is given by,
\begin{eqnarray}
\frac{1}{\tau^{esc}_{i,s}} = - \frac{1}{\int_{0}^{L_b} dx
 |\psi_{i,s}(x,t)|^2 }\frac{d }{dt}\int_{0}^{H} dx
 |\psi_{i,s}(x,t)|^2  \,.
 \label{eq:escape_rateIII}
\end{eqnarray}
Fig.~\ref{Fig:self_consistent_TDSE}d shows the escape rate from the
second quasi-bound state versus $k_m$. The escape rate peaks when
the ``effective velocities'' in the well and metal match:
$k_{m}$/$m_0$$\sim$$\pi$/$\{$$m_{sc}$($\ell_2$-$\ell_1$)$\}$.
Typical values of Fermi wave vectors in normal and ferromagnetic
metals fit to the right part of this figure where the escape rate
decreases with $k_m$. The escape rate from each of the bound states,
shown in Fig.~\ref{Fig:self_consistent_TDSE}b, into minority states
of Fe is nearly twice the escape rate into majority states. The spin
dependent current density due to escape of localized electrons is,
\begin{eqnarray}
J_{2D,s} \simeq q\sum_i \frac{\widetilde{n}_i}{2\tau^{esc}_{i,s}}\,,
 \label{eq:escape_rateI}
\end{eqnarray}
where $\widetilde{n}_i$ is the areal density of electrons in the
i$_{th}$ state whose energy is higher than the Fermi energy of the
ferromagnet. It is assumed that spin relaxation time in the well is
faster than the escape time ($n_{i,s}$$\simeq$$n_i$/2). The spin
relaxation time in the well is around tens of ps
\cite{Malinowski_PRB00} whereas the escape rate is $\sim$1~ns
\cite{footnote1}. Spin relaxation in the well does not cancel the
spin-polarization in the bulk region. This is due to the fast
spin-conserving capture process of free electrons by the well (e.g,
emitting longitudinal optical phonons or carrier-carrier scattering
with electrons of the degenerate well \cite{Deveaud_APL88}). This
means that an electron which escapes from the well into the
ferromagnet is replenished by an electron with the same spin from
the bulk region before spin relaxation takes place. The bulk region
is left with more spin-up (down) electrons if it provides the well
with more spin-down (up) electrons.

\begin{figure}
\includegraphics[height=6cm,width=8cm]{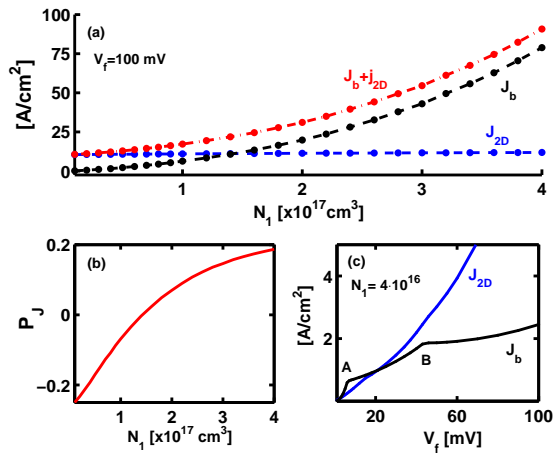}
\caption{(a) Extracted current density across the GaAs/Fe junction
versus the background doping. The black (blue) line denotes the
current due to free (localized) electrons. (b) Spin-polarity of the
total current shown by the red line in (a). (c) Current density
versus the bias across the junction in a case of low background
doping. In all cases V$_G$=0.} \label{fig:IV}
\end{figure}

Free electrons take part in tunneling when the semiconductor is not
depleted near the gate, i.e., V$_G$$\sim$0. The spin dependent
current density from tunneling of free electrons is given by
\cite{tunneling_phenomena}:
\begin{eqnarray}
J_{b,s} \!= \! \frac{4 \pi  m_{sc} q  }{h^3} \!\! \int_{0}^{E_{max}}
\!\!\!\!\!\!\!\!\!\! \!\!\! dE \big( f_{sc}(E)\!-\!\!f_{fm}(E) \big)
\!\!\! \int_{0}^{E} \!\!\!\!\!\! dE_{||}  T_s(\!E\!-\!E_{||}\!)\,,
 \label{eq:current_density}
\end{eqnarray}
where $f_{sc}$ and $f_{fm}$ are the Fermi distribution functions in
the semiconductor and ferromagnet, respectively. E is the total
kinetic energy taken from the semiconductor conduction band edge in
the bulk region. E$_{||}$ is the part of E due to the electron's
motion in parallel to the S/F interface. At low temperatures and in
forward bias E$_{max}$ is few meV above the semiconductor chemical
potential. T$_s$(E-E$_{||}$) is the spin dependent specular
transmission coefficient of the current. It is calculated by
applying the transfer-matrix method for the S/F potential. This
procedure includes the resonating behavior of free electrons due to
the well \cite{Stiles_PRB93}. Fig.~\ref{fig:IV}a shows the current
contributions from free and localized electrons versus the
background doping N$_1$ when V$_G$=0. The potential well includes
three localized states and is only mildly affected with changing
N$_1$ if N$_1$$<<$N$_2$. Therefore J$_{2D}$ increases only with 10
percent for the shown interval of N$_1$. On the other hand, J$_b$
strongly depends on N$_1$ via $E_{max}$$\propto$N$_1^{2/3}$ and
T$_s$(E-E$_{||}$)$\propto$exp(C$\cdot$N$_1^{2/3}$).
Fig.~\ref{fig:IV}b shows the spin polarity of the current,
P$_J$=(J$_{\uparrow}$-J$_{\downarrow}$)/J where J=J$_b$+J$_{2D}$ and
J$_s$=J$_{b,s}$+J$_{2D,s}$ (s=$\uparrow$,$\downarrow$). The critical
background doping for which P$_J$=0 is
$\sim$1.5$\cdot$10$^{17}$~cm$^{-3}$. This is not exactly the density
at which J$_b$=J$_{2D}$ because $|$P$_{J_b}$$|$ and
$|$P$_{J_{2D}}$$|$ are slightly different. For the spin-switch to
work, the background doping density must exceed this critical
density.

Our theory may also explain the spin-imaging measurements by Crooker
\textit{et al.} \cite{Crooker_Science05}. In this experiment, an
n-type GaAs/Fe structure with a background doping of
N$_1$=2$\cdot$10$^{16}$~cm$^{-3}$ was studied and the spin
polarization near the forward biased junction was opposite to the
one which is expected by tunneling of free electrons. The authors
have suggested that the reason for the opposite sign might be due to
spontaneous spin polarization caused by reflection of free electrons
\cite{Ciuti_PRL02}. However, we mention that spontaneous spin
polarization, caused by reflection of free electrons, changes its
sign if high-energy electrons are involved \cite{Ciuti_PRL02}, for
example by photo-excitation in the barrier region
\cite{Epstein_PRB02}. These conditions are not the case in
Ref.~\cite{Crooker_Science05} where the transport near the forward
biased junction is dominated by low-energy electrons whose energy
cannot exceed the chemical potential  by more than a few k$_{_B}$T.
In this case the sign of the spin either by reflection or
transmission calculations is opposite to the measured sign. The
results of Figs.~\ref{fig:IV}a and \ref{fig:IV}b show that including
the escape process of localized electrons explains the puzzling
measurement. Our analysis is also consistent with the longitudinal
optical phonon signature in the low temperature conductance
measurement of Fe/GaAs contacts by Hanbicki et \textit{al.}
\cite{Hanbicki_APL03}. The signature in the forward direction had
remained an open question and our study suggests that it is possibly
due to the capture process of electrons into the well.

To study the low background doping case, we show in
Fig.~\ref{fig:IV}c the current density versus the bias across the
S/F junction bias when N$_1$=4$\cdot$10$^{16}$. The chemical
potential, $\mu_{sc}$, lies about 6~meV above the conduction band
edge in the bulk region. When 0$<$V$_F$$<$$\mu_{sc}$, part of the
free electrons cannot tunnel due to Pauli-blocking from the
ferromagnet side. In this region, J$_b$ increases rapidly because
blocked electrons become available for tunneling with increasing the
bias. This behavior ceases when all of the free electrons can take
part in the current (feature A). On the other hand, for localized
electrons this behavior persists until the ground state energy is
above the Fermi energy of the ferromagnet
(0$<$V$_F$$\lesssim$100~mV). When V$_F$$\simeq$20~mV there is a
critical density of localized electrons beyond which
J$_{2D}$$>$J$_b$ and consequently the spin polarity changes sign
\cite{footnote2}. We also see a second feature (B) in J$_b$ when
V$_F$$\sim$45meV. In this case, the conduction band profile leads to
a relatively strong transmission of low-energy free electrons
(Ramsauer-Townsend resonance). The summation over the kinetic energy
in the parallel plane smears this peak in the I-V curve
(Eq.~(\ref{eq:current_density})).
%

Our free electron model neglects the full electronic band structure.
It was pointed out in a number of theoretical studies that the spin
injection is nearly perfect in ideal Fe/GaAs(001) structures
\cite{MacLaren_PRB99, Wunnicke_PRB02}. Of all d orbitals centered on
the Fe atoms, only the d$_{z^2}$ orbital leads to $\sigma$-type
overlap with the semiconductor states. Along the $\Gamma$-Z
tunneling direction and across the Fermi surface of bcc Fe this
corresponds to majority electrons of the $\Delta_{1}$ band states
with wave vector of about 1~\AA$^{-1}$. However, in real interfaces
disorder is inevitable and the spin polarization drops dramatically
\cite{Zwierzycki_PRB03}. For example, Fe substitutes at the top
As-terminated monolayer may lead to strong square in-planar bonding
via the d$_{xy}$ orbital with the semiconductor ligands
\cite{pauling_book}. This opens a transmission channel to minority
electrons of the $\Delta_{2'}$ band states with wave vector of about
0.5~\AA$^{-1}$. Our free electron modeling predicts that when the
bcc Fe is replaced by a zinc blende MnAs the accumulated spin at the
extracting region should have opposite sign to that of the simulated
Fe/GaAs case. This is due to the exchanged amplitudes of minority
and majority wave vectors \cite{Zhao_PRB02}.

In conclusion we have explained the nature of spin extraction from a
semiconductor into a ferromagnet. The inhomogeneous doping at the
semicondutor creates surface bands from which the preferred
extracted spin is opposite to that from the bulk conduction band. A
particular consequence is a proposed switch in which a non-magnetic
gate monitors the spin polarization in a semiconductor. The switch
utilizes a ferromagnet to filter either of the spin species
depending on the gate bias. The switch structure is closely related
with the double-gate CMOS technology. As such, back-gates may
replace current carrying wires on top of ferromagnetic contacts in
semiconductor spin-based logic circuits
\cite{dery_magnetic_computer}, thus enabling spintronics without
magnetic fields. Our study also predicts that the magneto resistance
effect in a spin valve structure should have opposite sign to the
preceding analysis.

This work is supported by NSF DMR-0325599.


\begin{thebibliography}{99}
\bibitem{footnote1} The total escape rate from the i$_{th}$ localized state may be
readily estimated using a WKB approximation,
$1/\tau^{esc}_{i,\uparrow}$+$1/\tau^{esc}_{i,\downarrow}$$\sim$$(\omega_0/2)$$\cdot$$exp\{2(V_F+E_i-\phi_b)/\hbar\omega_2\}$
where
$\omega_{0,2}^2$=(q$^2$N$_{0,2}$)/(m$_{sc}$$\epsilon_r$$\epsilon_0$)
with N$_0$ being the averaged doping in the well region. E$_i$ is
the energy of the bound state.
\bibitem{footnote2} We are indebted to Professor P. Crowell for communicating his experimental results
prior to their publication.
\end{thebibliography}
\end{document}